\documentclass[conference]{IEEEtran}
\IEEEoverridecommandlockouts
\usepackage{cite}
\usepackage{amsmath,amssymb,amsfonts}
\usepackage{algorithmic}
\usepackage{graphicx}
\usepackage{textcomp}
\usepackage{xcolor}
\usepackage{subcaption}
\usepackage{tabularx}
\usepackage{fancyvrb}
\usepackage{listings}
\makeatletter
\def\FV@BadEndError{%
  \@warning
      {Extraneous input  between
        \string\end{\FV@EnvironName} and line end}%
      \let\next\FV@EndScanning}
      
\def\BibTeX{{\rm B\kern-.05em{\sc i\kern-.025em b}\kern-.08em
    T\kern-.1667em\lower.7ex\hbox{E}\kern-.125emX}}
\begin{document}

\title{Automated Query Generation for Design Pattern Mining in Source Code\\
\thanks{This material is based upon work supported in part by the University of Washington Bothell, Computing \& Software Systems Division}
}

\author{
\IEEEauthorblockN{Jeffy Jahfar Poozhithara}
\IEEEauthorblockA{\textit{Computing \& Software Systems}\\
\textit{University of Washington Bothell}\\
Bothell, USA \\
jeffyj@uw.edu}
\and
\IEEEauthorblockN{Hazeline U. Asuncion}
\IEEEauthorblockA{\textit{Computing \& Software Systems}\\
\textit{University of Washington Bothell}\\
Bothell, USA \\
hazeline@uw.edu}
\and
\IEEEauthorblockN{Brent Lagesse}
\IEEEauthorblockA{\textit{Computing \& Software Systems}\\
	\textit{University of Washington Bothell}\\
	Bothell, USA \\
	lagesse@uw.edu}
}

\maketitle


\begin{abstract}
Identifying which design patterns already exist in source code can help maintenance engineers gain a better understanding of the source code and determine if new requirements can be satisfied.  There are current techniques for mining design patterns, but some of these techniques require tedious work of manually labeling training datasets, or manually specifying rules or queries for each pattern.  To address this challenge, we introduce Model2Mine, a technique for automatically generating SPARQL queries by parsing UML diagrams, ensuring that all constraints are appropriately addressed. We discuss the underlying architecture of Model2Mine and its functionalities. Our initial results indicate that Model2Mine can automatically generate queries for the three types of design patterns (i.e., creational, behavioral, structural), with a slight performance overhead compared to manually generated queries, and accuracy that is comparable, or perform better than, existing techniques. 
\end{abstract}


\begin{IEEEkeywords}
Design Pattern Mining, Security Design Pattern Mining, Semantic Web, RDF, SPARQL, UML Diagrams\end{IEEEkeywords}

\section{Introduction}

Design patterns are general purpose solutions to recurring software engineering problems. It has advantages such as enhancing re-usability and maintainability by furnishing an explicit specification of class and object interactions and their underlying intent \cite{b27}. Secure design patterns \cite{b28} are reusable components that not only addresses common vulnerabilities but also reduce the high cost and efforts associated with implementing security at a later stage \cite{b20}. Ever since they have been introduced, much research has gone into design patterns and secure design patterns, as they impact design, development, and maintenance stages of software engineering.

Since design patterns assist with satisfying requirements, it is important for maintenance engineers to determine which patterns are already present in the code.  Finding design patterns can be time-consuming, due to manual work required to reverse engineer the code \cite{b7}.  Meanwhile, other techniques also require manual work before design patterns could be mined.  This may involve the time-consuming task of manual labeling training data \cite{b13} or manual specification of patterns for mining (e.g., rules \cite{b45}, queries \cite{b6}).  Another key challenge in automation is the variations in implementations of design patterns making direct pattern matching infeasible.  This is especially true with secure design patterns, which have a higher level of variability than object-oriented design patterns \cite{b40} \cite{b41}.


To address these challenges, we developed Model2Mine which automatically generates queries from UML Class Diagrams \cite{b2} to mine design patterns.  We leverage Semantic Web technology, such as Resource Description Framework (RDF) \cite{b36}, CodeOntology \cite{b33} \cite{b34} and SPARQL \cite{b21} in developing this generator.  An RDF graph shows relationships between resources (which could be data) and these relationships are represented as triples \cite{b36}.  Code Ontology is used to convert source code to RDF triples, and it preserves all relationships between elements within code (e.g., between packages, classes, and methods  \cite{b33} \cite{b34}.  Once we have an RDF graph of the source code, we can can retrieve triples using SPARQL queries \cite{b21}.  Our technique, Model2Mine, automatically generates these SPARQL queries from an XML Metadata Interchange (XMI) \cite{b42} of a UML Class diagram.

Model2Mine is feasible to use because repositories of common design patterns have already been created \cite{b15} \cite{b27} and they already include UML Class diagrams in their description.  This also applies to security design patterns as many of them also include Class diagrams \cite{b43} \cite{b44}.  

In this paper, our main contribution is a language agnostic approach that is fully automated with the ability to account for implementation variants of any design pattern. Compared to other methods for design pattern mining in source code, the ease of use of this tool comes from the fact that it does not involve any manual training stage and does not require defining rules and queries. The second contribution is that Model2Mine incorporates behavioral aspects of a pattern in addition to structural characteristics by incorporating stereotypes and filters. Thirdly, the paper discusses the various ways in which accuracy can be enhanced when mining design patterns using SPARQL queries. 

We assessed Model2Mine using two types of evaluation.  First, we compared our automatically generated SPARQL queries against manually constructed queries, for different types of design patterns (i.e., creational, behavioral and structural patterns \cite{b27}).  The automatically generated SPARQL queries were comparable to manually constructed queries, with only slight differences in running time.  We also assessed the accuracy of mined queries.  Our results thus far indicate that they are also comparable, or perform better, than existing techniques \cite{b5}, \cite{b6}, \cite{b9},  \cite{b13}, \cite{b16}, \cite{b17}, \cite{b18}, \cite{b19}, \cite{b31}.

This paper is organized as follows.   We start with a motivation of our work. A comparison of Model2Mine with existing methodologies is discussed in Section~\ref{LiteratureReview}.  This is followed by a discussion of modeling and Semantic Web technologies used (Section~\ref{Background}).  Tool design, with a discussion of each module in the tool, is presented in Section~\ref{Design}. We then discuss our evaluations, limitations and challenges in Sections ~\ref{Validation} and~\ref{Discussion} respectively. 

\section{Motivation}
\label{Motivation}
Significant work has gone into identifying design patterns that can be used by software engineers \cite{b15} \cite{b27}, but these patterns assume that a developer is working on the design phase of software.  Many times, however, a maintenance engineer works on an existing codebase, and it is unclear which design patterns already exist in the source code.  

The ability to identify existing patterns in source code is especially important for legacy code that needs to meet new security requirements.  Model2Mine serves the purpose of helping security engineers to rapidly understand which, if any, existing security mechanisms (i.e., security design patterns) have been designed into the existing code base. The mined security design patterns can then be compared with security requirements.

\begin{figure}
	\centering
	\begin{subfigure}{.3\columnwidth}
		\centering
		\includegraphics[width=0.5\columnwidth]{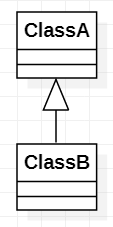}

		\label{fig:sub1}
	\end{subfigure}%
	\begin{subfigure}{.7\columnwidth}
		\centering
		\includegraphics[width=\columnwidth]{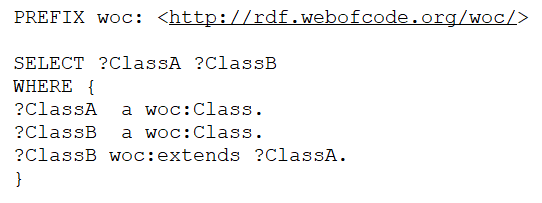}

		\label{fig:sub2}
	\end{subfigure}
	\caption{UML Diagram and corresponding SPARQL query for a simple inheritance relationship}
	\label{fig:simple_inheritance}
\end{figure}

\section{Background: Modeling \& Semantic Web}\label{Background}
In this section we provide background on the various technologies we use for Model2Mine.

\textbf{Modeling:} 
The Unified Modeling Language (UML) is a general-purpose modeling language intended to provide a standard way to visualize the design of a system \cite{b2}. A UML Class Diagram has components like Classes and Interfaces which in turn contains attributes, operations. Classes are connected using relationships including Generalization, Association, Composition, Collaboration and Interface Realization.


There are various UML editing tools that enables users to create UML Diagrams.  One of these tools is StarUML \cite{b49}. We use StarUML to create UML Class Diagrams of various design patterns (e.g., Proxy, Visitor, Factory, Builder).  We also used the StarUML XMI plugin to convert model (.mdj) and fragment (.mfj) files of these design patterns to XMI files.  These XMI files serve as input to Model2Mine.



\textbf{Modeling Metrics:} 
SDMetrics is an Object Oriented design quality measurement tool for UML \cite{b48}. SDMetrics analyzes the structure of UML models and works with all UML design tools that support XMI. Although the software is rich with features like comprehensive design measurements, automated design rule checks and an interactive UI, the only functionality we use in this project is the Open Core library used in its backend that parses UML Files stored as XMI. It supports all XMI versions currently in use.  It also has a flexible custom XMI import, configurable to support proprietary UML metamodel extensions and tools that deviate from XMI standards.



\textbf{Semantic Web:} 
Mine2Model is built on top of various Semantic Web technologies: RDF, CodeOntology, and SPARQL.
An RDF graph is a finite set of RDF triples \cite{b36}. RDF triples contain facts, which are relationships between resources.  Resources are represented as nodes, relationships are represented as edges.  The vocabulary for RDF graphs is three disjoint sets: a set of URIs $V_{uri}$, a set of bnode identifiers $V_{bnode}$, and a set of well-formed literals $V_{lit}$. The union of these sets is called the set of RDF terms. An RDF triple is a tuple $(s, p, o) \in ( V_{uri} \cup V_{bnode}) \cdot V_{uri} \cdot (V_{uri} \cup V_{bnode} \cup V_{lit})$ .

CodeOntology is a building block of the Web of Code, an attempt to leverage code in a semantic framework \cite{b33} \cite{b34}. The CodeOntology has an exposed API to parse source code of OpenJDK8 as well as result set of parsing open source code on Github through the GitHub API.  Its framework is composed of three main components: Ontology, Parser and Datasets. 

The ontology component is designed to model the domain of object-oriented programming languages. It is written in OWL 2 and is mainly focused towards the Java programming language, but it can be replaced to represent more languages. The modelling process underlying the creation of the ontology has been guided by common competency questions that usually arise during software processes and has been inspired by a re-engineering of the Java abstract syntax tree.  

The parser component analyzes Java code to serialize it into RDF triples. Internally, the RDF triple extraction is managed by a Spoon \cite{b47} processor invoked for every package in the input project. The RDF serialization process is handled using Apache Jena \cite{b46}. It is able to extract structural information common to all object-oriented programming languages, like class hierarchy, methods and constructors. Optionally, it can also serialize into RDF triples all the statements and expressions, thereby providing a complete RDF-ization of source code. The RDF serialization of a Java project acts in three steps.  First the project is analyzed to download all of its dependencies and load them in class path.  Then an abstract syntax tree of the source code and its dependencies is built and processed to extract a set of RDF triples.

We also use SPARQL.  A building block for SPARQL queries is Basic Graph Patterns (BGP). A SPARQL BGP is a set of triple patterns. A triple pattern is an RDF triple in which zero or more variables might appear. Variables are taken from the infinite set $V_{var}$ which is disjoint from the above-mentioned sets \cite{b1}. 

SPARQL is a query language and a protocol for accessing RDF graphs \cite{b1}. SPARQL takes the description of what the application wants, in the form of a query, and returns that information, in the form of a set of bindings or an RDF graph. 

A sample SPARQL query that searches for all entries that has the name attribute set to value Smith is as follows:\\

\newsavebox\myfoaf
\begin{lrbox}{\myfoaf} 
\begin{minipage}{\columnwidth} 
\begin{Verbatim}[numbers=left,xleftmargin=5mm] 
PREFIX foaf: <http://xmlns.com/foaf/0.1/>
SELECT ?name
WHERE {  
?person foaf:name “Smith” .
}
\end{Verbatim}	 
\end{minipage} 
\end{lrbox} 
\resizebox{0.8\columnwidth}{!}{\usebox\myfoaf} \\

\begin{figure*}
	\centering
	\includegraphics{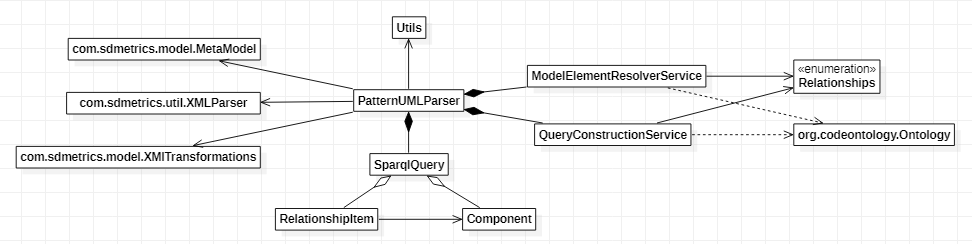}
	\caption{Object-oriented Design of the UML to SPARQL Converter Library}
	\label{fig:architecture}
\end{figure*}

The query has PREFIX, SELECT and WHERE sections where PREFIX defines the database schema being queried from, SELECT statement defines the attributes being extracted and WHERE statement defines the various constraints that need to be matched to extract the results. The WHERE statement can have one or more constraints including a FILTER statement.


\section{Literature Review}\label{LiteratureReview}

Earlier approaches for detecting design patterns in source code ranged from sub-graph matching \cite{b16}, \cite{b17}, \cite{b19}  and ontology based techniques \cite{b6}, \cite{b9} to using machine learning techniques \cite{b5}, \cite{b13},\cite{b18}, \cite{b31} and sequence diagrams \cite{b9}. A detailed meta-analysis of various design pattern mining approaches is discussed in \cite{b14}. 

The construction RDF triples from UML diagrams is discussed in \cite{b39}. Our technique introduces a novel method and tool called Model2Mine, to generate SPARQL queries automatically by parsing UML diagrams. Model2Mine uses semantic web based technologies to convert source code to RDF triples and to query the triples using SPARQL queries. This technique not only removes the bottleneck of manually constructing queries but also enables bulk parsing of projects and creating datasets for source code mining research. 

There are ontology-based approaches to mining design patterns (e.g., \cite{b6}, \cite{b9}).  One approach uses Semantic Web technologies for automatically detecting design patterns \cite{b6}.  However, this requires manual specification of queries and rules. That is, the SPARQL queries had to be manually constructed for each pattern intended to be mined. Their queries handle implementation variations using Union operations by defining each component and associated relationships within the Union Operation block. However, since the SELECT statement and component declaration is common, this only incorporates implementation variants that have exactly same number of target components. Another technique uses a knowledge base and inference rules to detect the design patterns that are similar in structure  \cite{b9}. The target system design, including a class diagram and its associated sequence diagrams, are analyzed and translated into knowledge concepts in ontology in terms of RDF/OWL elements. The detection is performed by semantically searching their predefined knowledge base of the expected design patterns and their corresponding detecting inference rules through SWRL and SQWRL. Our method uses a similar approach that relies on ontology by converting source code to RDF triples. However, we not only mine for structurally similar patterns, but also addresses behavioral and creational patterns as well. We achieve this by using SPARQL queries. 



Numerous researchers have identified language specific solutions to design pattern mining in object oriented languages like C++ and Java that includes both manual \cite{b12} and automated techniques \cite{b26} \cite{b37}. The underlying idea of creating a language-agnostic parser is similar to the multi-stage filtering strategy in \cite{b25} as they also address the behavioral and creational patterns in addition to filtering structural similarity. However their extractor was developed only for C++ and the Abstract Object Language (AOL) representations of each pattern had to be constructed manually.

The IDEA (Interactive DEsign Assistant) system is another tool that matches a UML diagram of a design pattern against a class being implemented by verifying if the implementation can be improved to match the design pattern \cite{b24}. However, this only supports verification of one class at a time in the source code due to scalability issues. Applying on a distributed set of open source projects is difficult. 

Other techniques use source code metrics and machine learning to detect patterns without using strict structural constraints to cater to variations in implementation of the patterns in different projects such that minor variation in structure will not lead to false negative results \cite{b5} \cite{b18} \cite{b13}. However, this lack of strict constraints leads to a large number of false positive results. Finally, it also requires tedious manual training for each pattern that needs to be detected. The similarity score comparison of graph vertices used in \cite{b31} has the same limitation. The advantage of our model is that it caters to accommodating variations in implementation without compromising on accuracy and also removing the requirement for manual training for each pattern. 

There also have been approaches to detect patterns from software documentations \cite{b30}. However they require the descriptive and prescriptive architecture to be the same for the model to perform accurately. 

\section{Design}\label{Design}

As we mentioned, Model2Mine is built on top of the semantic web technology discussed in Section~\ref{Background}. It uses the XMI representations of UML Class diagrams to generate queries for mining these patterns in source code, which is represented as an RDF graph.  An object-oriented design of the tool is as shown in Figure~\ref{fig:architecture}. 

The architecture follows a modular design with separation of concerns. For instance, the task of identifying components and relationships are handled by ModelElementResolverService. On the other hand, the task of constructing the query from identified Components and RelationshipItems is handled by QueryConstructionService. The two services are completely decoupled. Model2Mine was designed to enhance the re-usability and portability of the individual modules. Model2Mine can be extended to support more relationship types and component types with minimal changes.  

Implemented objects are described in detail below.

\subsection{PatternUMLParser}
The PatternUMLParser class contains the parseXMIFile method and saveOutputAsText method that parses UML Class diagrams in .xmi files and saves SPARQL queries as .rq files respectively. A Model object is created using the SDMetrics Open Core library using the XMI file. Once the diagram is parsed as Java Model Object, each ModelElement in the model is iteratively converted into a Component or RelationshipItem object. Further, the library iteratively analyzes each component and relationship in the diagram. It creates a SPARQL query which includes two parts: A SELECT statement and a WHERE clause. The query is a string formed by concatenating each RDF triple generated by analyzing relationships (e.g., associations, interface realizations, generalizations) in the Class diagram as well as constraints like data type and visibility.

\begin{figure}
	\centering
	\includegraphics[width=\columnwidth]{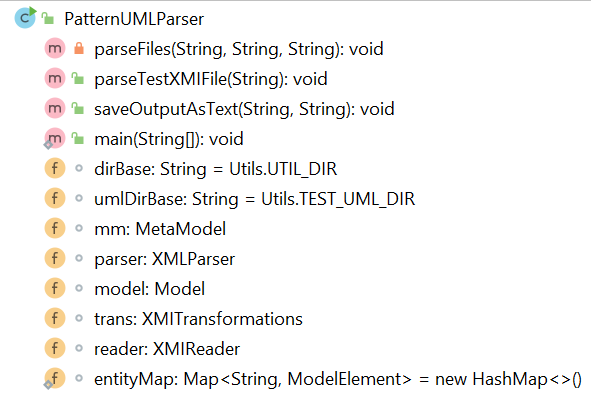}
	\caption{Hierarchical map of Model Elements created by parsing XMI using SDMetrics Open Core library}
	\label{fig:patternUmlParser}
\end{figure}

The PatternUMLParser relies on ModelElementResolverService to resolve whether the ModelElement being parsed is relevant for constructing SPARQL query or not. The ModelElementResolverService constructs a blank SparqlQuery object and each element is added to the list of Components or RelationshipItems in the SparqlQuery object being constructed. Once all elements are checked, PatternUMLParser uses the QueryConstructionService to construct the query attribute of the SparqlQuery object. The Model object that contains hierarchical map of ModelElements constructed by parsing XMI representation of a UML diagram is shown in Figure~\ref{fig:patternUmlParser}. The class parses XMI file to identify UML elements defined in the MetaModel object based on the element to XMI keyword mapping defined in the XMITransformations object. After the construction of SPARQL query as explained above, the output is saved as an .rq file in the path defined in the Model2Mine util files.

\subsection{ModelElementResolverService}
This service resolves whether a ModelElement should be appended into the query or not and in what format. Dedicated methods resolving all types of relationships are defined here and relies on the Enumeration (enum) Relationships to resolve relevant relationships. 

\subsection{Component}
A component are the nodes in a UML diagram between which relationships exist. Components could be classes, interfaces, methods or other attributes. 

\subsection{Relationships}
This is an enumeration identifying all the relationships that appear between two components in a SPARQL query generated from a UML Diagram. This contains both relations between two classes, between a class and its attributes and methods, and between methods and its parameters. That is, in addition to the relationships like Generalization, Interface Realization, Association and Composition, this has entries corresponding to generation of triples with relations such as woc:hasReturnType, woc:hasMethod, woc:hasParameter etc. This is maintained as a separate Enumeration so that construction of constraint triples can be generalized as a RelationshipItem explained in section~\ref{RelationshipItem} where each element of the triple has types:
(Component, Relationships , Component).

\subsection{RelationshipItem}\label{RelationshipItem}
A relationshipItem has three attributes: a $fromItem$, a $toItem$, and a $relationshipType$ which has a value from the Relationships enumeration describing the relationship $fromItem$ has to the $toItem$. Both $fromItem$ and $toItem$ are of Component type. An RDF triple can be constructed as
	
$ (fromItem,relationshipType,toItem)$

\subsection{QueryConstructionService}
Once the lists of components and relationships are constructed for a Model in consideration, the QueryConstructionService builds the SparqlQuery. Each component is added to the Select statement. Each component is also added in the WHERE clause defining it’s type. For example, if a Method is encountered, an RDF triple defining the element has woc:Method type is added.\\

\newsavebox\mya
\begin{lrbox}{\mya}
    \begin{minipage}{\columnwidth}
        \begin{Verbatim}[numbers=left,xleftmargin=5mm]
SELECT ?ClassA ?OperationA 
WHERE {
?ClassA  a woc:Class .
?OperationA  a woc:Method .
        \end{Verbatim}	
    \end{minipage}
\end{lrbox}
\resizebox{0.8\columnwidth}{!}{\usebox\mya} \\
	
After the components are all checked, we iterate on RelationshipItems.   An RDF triple is created based on what relation the fromItem has to the toItem is appended to the query. In the following snippet, ClassA is the $fromItem$, OperationA is the $toItem$ and woc:hasMethod is the $relationshipType$. \\

\newsavebox\myb
\begin{lrbox}{\myb}
	\begin{minipage}{\columnwidth}
		\begin{Verbatim}[numbers=left,xleftmargin=5mm]
SELECT ?ClassA ?OperationA 
WHERE {
?ClassA  a woc:Class .
?OperationA  a woc:Method .
?ClassA woc:hasMethod ?OperationA .
        \end{Verbatim}	
	\end{minipage}
\end{lrbox}
\resizebox{0.8\columnwidth}{!}{\usebox\myb} \\
	
		
Once all relationships are checked, the where clause is closed using a closing bracket “\}”. 


\subsection{SparqlQuery}
The SparqlQuery object has three attributes: a query string (which is the final SPARQL query output), a list of components used to construct the select statement,  and a list of relationship items used in the WHERE clause. 



\subsection{Ontology}
\begin{figure}
	\centering
	\includegraphics[width=\columnwidth]{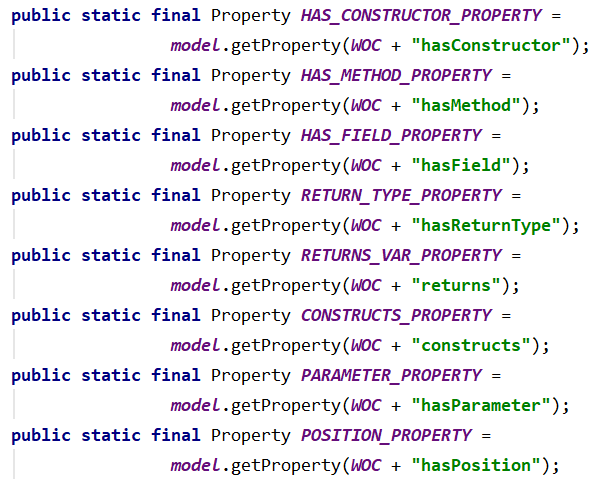}
	\caption{The dictionary available in Ontology class for all WOC entities and relations}
	\label{fig:Ontology}
\end{figure}

The Ontology.java file from the CodeOntology parser library defines a dictionary of Java String variables that has values from the Ontology library.  It defines different keywords required to construct RDF triples for source code mining. A snippet from the Ontology file provided as part of CodeOntology Parser showing various entities and relationships defined in the Ontology is shown in Figure~\ref{fig:Ontology}. It can be seen that relationships and entities are defined as static variables in this file. Model2Mine uses these constants instead of hard-coding Ontology keywords in its implementation. 

\section{Enhancements to Improve Accuracy}
When parsing a UML diagram to generate a query, the parser is required to make the component names in the query unique. For example, in a Visitor Design Pattern, both the interface Visitor and each of its implementations will have visit methods of the same name. Although the names are the same, the entities have independent existence and the RDF triples need to be distinguished. For example, in the following snippet, both Visitor and ConcreteVisitor have a VisitElementA method where ConcreteVisitor’s VisitElementA method overrides Visitor’s method during implementation. However, this snippet assumes both VisitElementA methods to be the same. \\

\newsavebox\myv

\begin{lrbox}{\myv}
	\begin{minipage}{\columnwidth}
		\begin{Verbatim}[numbers=left,xleftmargin=5mm]
Visitor a woc:Interface .
ConcreteVisitorA a woc:Class .
ConcreteVisitor woc:Implements Visitor .
Visitor woc:hasMethod VisitElementA .
ConcreteVisitor woc:hasMethod VisitElementA .
		\end{Verbatim}
	\end{minipage}
\end{lrbox}

\resizebox{0.8\columnwidth}{!}{\usebox\myv} \\

The correct way of representing this scenario however, is \\

\newsavebox\myva
\begin{lrbox}{\myva}
	\begin{minipage}{\columnwidth}
		\begin{Verbatim}[numbers=left,xleftmargin=5mm]
Visitor a woc:Interface .
ConcreteVisitorA a woc:Class .
ConcreteVisitor woc:Implements Visitor .
Visitor woc:hasMethod VisitElementA1 .
ConcreteVisitor woc:hasMethod VisitElementA2 .
VisitElementA1 woc:overrides VisitElementA2 .
		\end{Verbatim}
\end{minipage}
\end{lrbox}

\resizebox{0.8\columnwidth}{!}{\usebox\myva}\\

\bgroup
\def\arraystretch{1.5}%
\begin{table*}
	\centering
	\begin{tabular}{|l|l|l|l|l|l|}
		\hline
		\textbf{Components}   & \textbf{Qualifiers}     & \textbf{Visibility} & \textbf{Modifiers} & \textbf{Relationships }        & \textbf{Stereotypes}    \\ \hline
		Classes      & hasMethod      & Public     & Static    & Association           & hasConstructor \\ 
		Methods      & hasType        & Private    & Abstract  & Generalization        & overrides      \\ 
		Constructors & hasReturnType  & Protected  & Final     & Aggregation           &                \\ 
		Fields       & hasModifiers   &            &           & Composition           &                \\ 
		Method       & hasField       &            &           & Interface Realization &                \\ 
		Parameters   & hasParameter   &            &           & Dependency            &                \\ 
		Interfaces   & hasConstructor &            &           &                       &       \\ \hline        	
	\end{tabular}
	\caption{Feature Coverage of the Library}
	\label{table:coverage}
\end{table*}
\egroup

An approach to make this possible would be to use uniquely generated identifiers for each component. However, when analyzing results, random generated identifiers would be difficult to interpret. Hence, in our library, the runningId of each element was appended to the Component’s name. Running ID is a sequence number that is auto generated when the XMI is being parsed using SDMetrics library.

For patterns that have multiple components of the same type, the query might assume the same component to be suitable for both the items in the SELECT statement. For example, for the SPARQL query given below that looks for a class with two methods, the following triples will also identify a class that has only one method as a result by substituting the same method for both MethodA and MethodB.\\

\newsavebox\myvfilter
\begin{lrbox}{\myvfilter}
	\begin{minipage}{\columnwidth}
		\begin{Verbatim}[numbers=left,xleftmargin=5mm]
ClassA a woc:Class .
MethodA a woc:Method .
MethodB a woc:Method .
ClassA woc:hasMethod MethodA .
ClassA woc:hasMethod MethodB .
		\end{Verbatim}
\end{minipage}
\end{lrbox}

\resizebox{0.8\columnwidth}{!}{\usebox\myvfilter}\\

In order to avoid this, the SELECT DISTINCT statement or FILTER statement has to be specified as shown below:\\

\newsavebox\myvfilterA
\begin{lrbox}{\myvfilterA}
	\begin{minipage}{\columnwidth}
		\begin{Verbatim}[numbers=left,xleftmargin=5mm]
ClassA a woc:Class .
MethodA a woc:Method .
MethodB a woc:Method .
ClassA woc:hasMethod MethodA .
ClassA woc:hasMethod MethodB .
FILTER(MethodA != MethodB)
		\end{Verbatim}
\end{minipage}
\end{lrbox}

\resizebox{0.8\columnwidth}{!}{\usebox\myvfilterA}\\

\begin{figure}
	\includegraphics[width=\columnwidth]{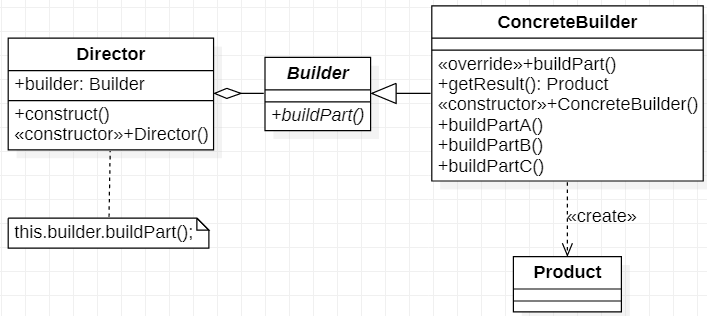}
	\caption{UML Diagram of Builder Pattern with stereotypes}
	\label{fig:stereotype}
\end{figure}

Another feature implemented to improve accuracy was the use of stereotypes. Although not part of the standard UML specification, numerous stereotypes have become popular among software engineers to differentiate or represent features like Constructors, Getters, Setters and Overriding of methods. When stereotypes are enabled, Constructors are differentiated from other Methods and the triples are constructed with woc:Constructor instead of woc:Method type and woc:hasConstructor instead of woc:hasMethod relationship. Similarly, properties of a child class that overrides properties of a parent class are differentiated with woc:overrides relationship. An example of Builder pattern with stereotypes enabled is shown in Figure~\ref{fig:stereotype}. This was observed to significantly reduce false positive results. This can be a powerful feature in improving accuracy of detection of patterns like Proxy. For example, when stereotypes are not enabled, the relationship between Client, Real Subject and Proxy classes are represented as \\
\newsavebox\myvproxy

\begin{lrbox}{\myvproxy}
	\begin{minipage}{\columnwidth}
		\begin{Verbatim}[numbers=left,xleftmargin=5mm]
?RealSubject woc:references ?Proxy8 .
?Client woc:references ?Subject .
\end{Verbatim}
\end{minipage}
\end{lrbox}
\resizebox{0.8\columnwidth}{!}{\usebox\myvproxy}\\

However, with stereotypes enabled, the scenario can be made more descriptive as:\\

\newsavebox\myvb
\begin{lrbox}{\myvb}
	\begin{minipage}{\columnwidth}
		\begin{Verbatim}[numbers=left,xleftmargin=5mm]
?proxyConstructor a woc:Constructor .
?Proxy woc:hasConstructor ?proxyConstructor .
?subjectConstructor a woc:Constructor .
?RealSubject woc:hasConstructor ?subjectConstructor .
?someMethod a woc:Method .
?Client woc:hasMethod ?someMethod .
?someMethod woc:references ?Proxy .
?someMethod woc:references ?RealSubject .
?ProxyRequestMethod woc:references ?RealSubject .
?someMethod woc:references ?subjectConstructor .
?someMethod woc:references ?proxyConstructor .
		\end{Verbatim}
	\end{minipage}
\end{lrbox}

\resizebox{0.75\columnwidth}{!}{\usebox\myvb} \\

Model2Mine exposes options whereby settings like suppressing visibility constraints and parsing stereotypes can be configured on a case to case basis. When visibility constraints are suppressed, constraint triples that states woc:hasModifier Public/Protected/Private are not generated. Similarly, the model looks for stereotypes only if the configuration is set true for parsing stereotypes. 

Including and excluding visibility constraints when generating query has to be decided on a case to case basis. Although Model2Mine has the flexibility to configure whether to include or exclude visibility constraints in the current execution, the user has to decide what setting is required for best performance of the current pattern in consideration. For instance, including visibility constraints ensure detection of wrongly implemented patterns, especially for creational patterns. This could be useful for code quality purposes, in cases when the work of a developer needs to be checked. For example, when implementing a Singleton pattern, if the unique Instance has public visibility instead of private, the implementation does not ensure instantiation is done only through the getInstance() method. That is, some class could wrongly access the uniqueInstance before its initialization. On the other hand, if visibility constraints are not included, a larger variation in implementation could be incorporated as this would compensate for the differences in coding style of different developers. For example, some developers might add access modifiers with each method even if it is automatically inherited from the parent class visibility while others might specify visibility only if it is different from the parent class. In some instances, visibility might only be a loose criteria unlike other hard structural constraints. In such scenarios, developers might have used a different visibility compared to the traditional implementation of a pattern as per documentation. In such cases, suppressing visibility might reduce the number of false negative results.

Other improvements in accuracy involve enabling detection of Static and Final modifiers from XMI representations which was not supported by the default Meta Models and XMI transformations distributed with the SDMetrics Open Core library. 

The coverage of constraints/relationships by the queries generated by this library are summarized in Table~\ref{table:coverage}.

\begin{figure*}
	\centering
	\includegraphics[width=\textwidth]{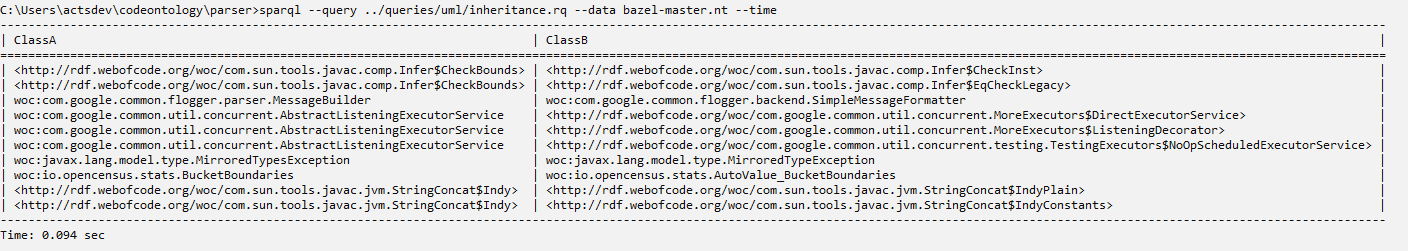}
	\caption{Result of parsing simple inheritance relationship using CodeOntology}
	\label{fig:inheritance_nt}
\end{figure*} 

\begin{figure}
	\centering
	\includegraphics[width=\columnwidth]{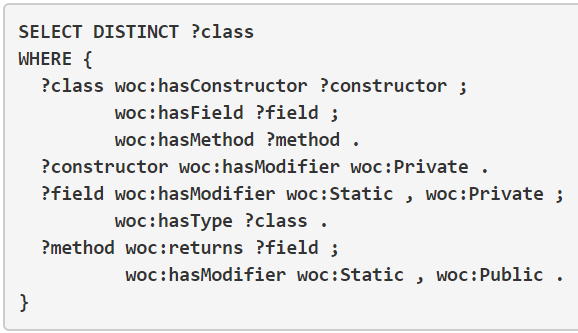}
	\caption{A manually constructed SPARQL query example for mining Singleton pattern on CodeOntology}
	\label{fig:manual_singleton_sparql}
\end{figure}

\section{Validation}\label{Validation}
We validated our library through the following: Feature Coverage and Quantitative Measures (which include accuracy and performance).

Model2Mine was used to generate SPARQL queries for representatives of each type of object-oriented design patterns: Creational Patterns (Abstract Factory, Builder, Singleton), Structural Patterns (Proxy, Adapter), Behavioral Patterns (Visitor, Strategy).

The queries were used to parse open source projects. The projects were first converted to .nt files with RDF triples using CodeOntology. The result of parsing SPARQL query for the simple inheritance relationship shown in Figure~\ref{fig:simple_inheritance} on Bazel \cite{bazel} project code limiting result to first 10 rows is shown in Figure~\ref{fig:inheritance_nt}. The execution output of a SPARQL query shows the entity from the project source code identified for each component in the SELECT statement of the query. In this example, the two entities in SELECT statement are ClassA and ClassB. Each entity is separated by $|$ symbol. Further, by passing the --time argument to SPARQL, it is possible to retrieve the total time taken to parse all the triples of the project for the current query. 

Dataset:  These two evaluations used source code from three open source projects. These were small projects, which contain 1741 \cite{iluwatar}, 2059 \cite{refactoringGuru} and 2034 \cite{jameszbl}, lines of code. These projects were chosen so that the same projects can be used for all the patterns we were analyzing as part of this study. While there are a number of open source projects that use one or the other design pattern, there are only few that contains all the three types (creational, structural and behavioral) of patterns. These projects were transformed into RDF triples without including dependency jars.

\subsection{Feature Coverage}

In order to compare if the constraints are properly incorporated, we compared manually constructed SPARQL queries with the generated queries for Builder, Factory and Singleton patterns. The results retrieved by the manual and generated queries on various open source projects were also compared. The manually constructed SPARQL query for singleton pattern is shown in Figure~\ref{fig:manual_singleton_sparql}. The corresponding query generated by Model2Mine is given below:\\

\newsavebox\myvsingleton
\begin{lrbox}{\myvsingleton}
\begin{minipage}{\columnwidth}
\begin{Verbatim}[numbers=left,xleftmargin=5mm]
PREFIX woc: <http://rdf.webofcode.org/woc/>

SELECT ?Instance ?SingletonOperation ?Singleton 
?uniqueInstance 
WHERE {
?Instance  a woc:Method .
?Instance woc:hasModifier woc:Public .
?SingletonOperation  a woc:Constructor .
?SingletonOperation woc:hasModifier woc:Private .
?Singleton  a woc:Class .
?Singleton woc:hasModifier woc:Public .
?uniqueInstance  a woc:Field .
?uniqueInstance woc:hasModifier woc:Private .
?Singleton woc:hasMethod ?Instance .
?Instance woc:hasReturnType ?Singleton .
?Singleton woc:hasConstructor ?SingletonOperation .
?Singleton woc:hasField ?uniqueInstance .
}
\end{Verbatim}
\end{minipage}
\end{lrbox}

\resizebox{0.75\columnwidth}{!}{\usebox\myvsingleton} \\

 Manually generated query uses shorthand notations and nested constraints corresponding to the same variable under one triple.  The automatically generated query does not use this shorthand notation due to the generic and iterative nature of the algorithm used in Model2Mine.  However, despite their differences, the coverage of constraints and corresponding results after running the query was similar (see Table~\ref{tab:manualvsgenerated}). The Nodes identified are the components that were part of the SELECT statement of the queries. Modifier Constraints are constraints related to non-access modifiers such as Static, Final and Abstract. Visibility constraints are triples related to access modifiers like Public, Protected and Private. Relationships column represents the relationships that were captured between different classes in the UML diagram. Execution time captures the average time taken for the SPARQL query of each type to execute over the 3 projects used in evaluation.

It was observed that the iterative algorithm of Model2Mine covered all the relevant components and relationships of the pattern including behaviors captured using stereotypes. The output obtained by parsing both manually constructed query and the generated query have exactly the same accuracy (f1-score of 0.9247). The execution time was slightly higher for the automatically generated query as SPARQL was required to reorder the constraints of the parsed query according to its internal relational algebra. The manually constructed queries were already arranged in the order that SPARQL expects it to be. This was verified by passing the --debug argument to SPARQL when parsing the generated queries over RDF triples of source code. 

\bgroup
\def\arraystretch{1.5}%
\begin{table}
	\centering
	\begin{tabular}{|l|l|l|l|}
		\hline
		\textbf{Pattern Type}   & \textbf{Pattern Name}     & \textbf{Precision} & \textbf{Recall}     \\ 
		\hline
         Creational  & Factory   & 100\%                             & 100\%                           \\
         & Singleton & 100\%                              & 86\%                      \\
         Behavioral & Visitor   & 100\%                              & 67\%              \\
         Structural  & Proxy     & 43\%                 & 75\%                      \\
        \hline        	
	\end{tabular}
	\caption{Precision and Recall observed for each pattern}
	\label{tab:accuracy}
\end{table}
\egroup

\bgroup
\def\arraystretch{1.5}%
\begin{table*}
	\centering
	\begin{tabular}{|p{0.15\columnwidth}|p{0.15\columnwidth}|p{0.2\columnwidth}|p{0.25\columnwidth}|p{0.25\columnwidth}|p{0.45\columnwidth}|p{0.15\columnwidth}|}
		\hline
\textbf{Pattern} &    & \textbf{Nodes Identified} & \textbf{Modifier Constraints} & \textbf{Visibility Constraints} & \textbf{Relationships} & \textbf{Execution Time (ms)} \\
 \hline
Singleton & Manual    & 4                & \checkmark                  & \checkmark                     & Extends, dependency, references, aggregation   & 125\\
& Generated & 4                & \checkmark       & \checkmark                    & Extends, dependency, references, aggregation & 152\\
\hline
Builder & Manual    & 4                & \checkmark                  & \checkmark                     & Extends, dependency, association, aggregation    & 94\\
& Generated & 8 (director class included)                & \checkmark       & \checkmark                    & Extends, dependency, association, aggregation & 157\\
\hline
Factory & Manual    & 4                & \checkmark                  & \checkmark                     & Extends     & 110\\
& Generated & 6                & \checkmark       & \checkmark             & Extends  & 131\\
		\hline        	
	\end{tabular}
	\caption{Comparison of Manual and Generated Query coverage}
	\label{tab:manualvsgenerated}
\end{table*}
\egroup

\subsection{Quantitative Measures} 
The precision and recall observed for the three projects that did not include dependencies are summarized in Table~\ref{tab:accuracy}. An average precision of 85.71\% and average recall of 82.29\% was achieved when parsing the three open source projects in Java. It was observed that, for patterns that can be better identified using unique stereotypes, precision and recall can be improved up to 100\%. In order to verify if the accuracy can be maintained for even larger code bases, the query was further executed on \cite{bazel}. 

\section{Lessons Learned}\label{Lessons}

\textbf{Handling Large Projects:} The tool was also evaluated over large projects like Bazel \cite{bazel} and JLibs \cite{jlibs} with 1965128 and 107075 lines of code respectively. For such large projects, triples were generated only if dependencies jars were included. Most large projects also have automated test cases like JUnit tests. This led to creation of .nt files with sizes larger than 1GB. In such projects, executing SPARQL queries required mentioning LIMIT statement in the query to limit the result to a specific number of rows due to system memory limitations and high execution time. Often times, executing SPARQL query takes 10+ hours to run without limit to the results. Such projects also poses challenge in manual verification of RDF triple creation of source code, as traditional text editors have file size limits of 1GB. In such scenarios, calculating false positive and false negative rates is not feasible. Further, results obtained with LIMIT statement specified were predominantly triples from the dependency jars which are difficult to manually verify unless decompiled.\\


\bgroup
\def\arraystretch{1.5}%
\begin{table*}
	\centering
	\begin{tabular}{|p{0.2\columnwidth}|p{0.2\columnwidth}|p{0.7\columnwidth}|p{0.7\columnwidth}|}
		\hline
		\textbf{Pattern Type}   & \textbf{Pattern Name}     &\textbf{Conditions for Best performance }   & \textbf{Limitations}   \\ 
		\hline
	Creational  & Factory   & Requires relaxing visibility constraints due to variation in developer practices  & Allowing single implementation scenarios lead to significant lose of precision    \\
	& Singleton & Stereotypes need to be incorporated to identify Constructor and unique Instance Getter method & Relaxing visibility constraints could lead to false positive detection of erroneous implementations \\
	Behavioral & Visitor   & Parameter type and return type constraints necessary    & Single implementation scenarios lead to low precision \\
	Structural  & Proxy     & Stereotypes need to be incorporated to identify Constructor and reference of constructors & Difficult to distinguish from visitor pattern leading to high False Positive Rate       \\                                      
		\hline        	
	\end{tabular}
	\caption{Criteria for best performance and limitations of each query}
	\label{tab:criteria}
\end{table*}
\egroup

\textbf{Identifying Best Conditions:} The performance of queries generated under different criteria were compared. Queries were generated by passing different configurations to PatternUMLParser like 1. Include Visibility 2. Suppress Visibility and 3. Include stereotypes. UML Diagram variants to include only interfaces/abstract classes with at least 2 implementations/extensions respectively (where applicable) was also considered. Query generated for each of these 4 variations of the patterns were executed over the 3 projects being evaluated to see how the False Positive, False Negative and True Positive rates varied. The outputs were compared to find the configuration that had the best performance in terms of precision and recall. The criteria for best performance and limitations observed for generated queries of each type was documented as shown in Table~\ref{tab:criteria}. 

\begin{figure}
	\centering
	\includegraphics[width=\columnwidth]{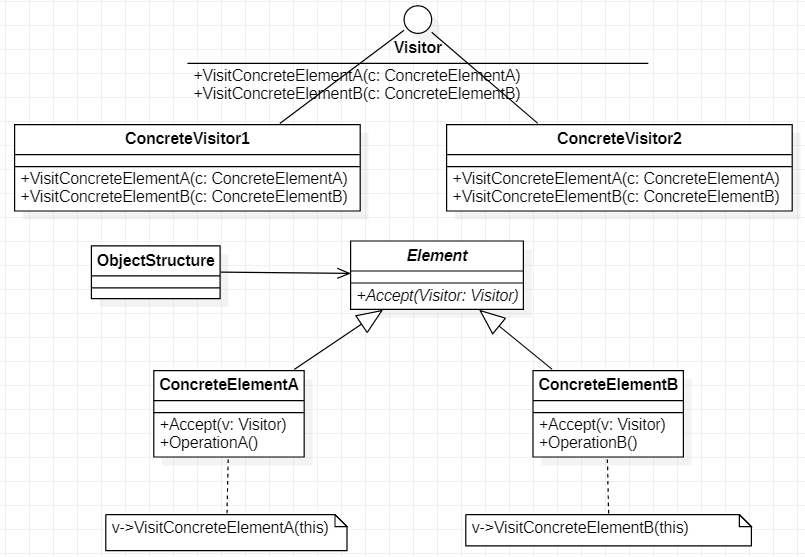}
	\caption{UML Representation of Visitor Design Pattern}
	\label{fig:visitor_uml}
\end{figure}

\begin{figure}
	\includegraphics[width=\columnwidth]{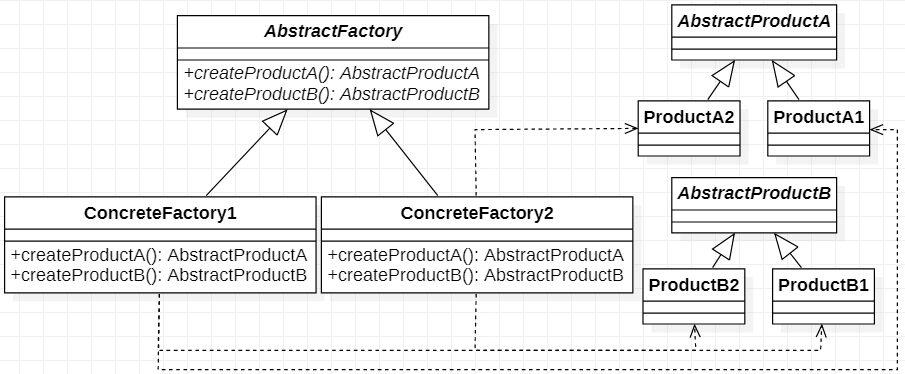}
	\caption{Variant of Abstract Factory that uses Abstract Classes}
	\label{fig:abstract_factory1}
\end{figure}

\begin{figure}
	\includegraphics[width=\columnwidth]{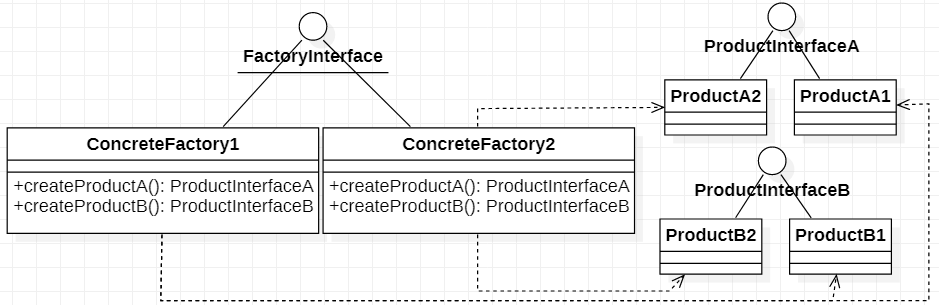}
	\caption{Variant of Abstract Factory that uses interfaces}
	\label{fig:abstract_factory2}
\end{figure}

\textbf{Handling Design Pattern Variations:} To handle design pattern variations, we will look at a design pattern that has variations in implementation, the Visitor Design Pattern (see  Figure~\ref{fig:visitor_uml}).  Different variations in developer implementations of these patterns were used to assess the ease in incorporating implementation variants. For example, two most common variant of Abstract Factory pattern are shown in Figure~\ref{fig:abstract_factory1} and Figure~\ref{fig:abstract_factory2}.

Model2Mine is able to generate queries for complex nested patterns. It does not limit the number of entities or relationships to be queried. The query will have the same granularity as the input UML diagram. The query generated by parsing the diagram shown in Figure~\ref{fig:visitor_uml} is as follows:\\

\newsavebox\myvc
\begin{lrbox}{\myvc}
	\begin{minipage}{\columnwidth}
		\begin{Verbatim}[numbers=left,xleftmargin=5mm]
PREFIX woc: <http://rdf.webofcode.org/woc/>

SELECT ?Visitor27 ?VisitConcreteElementA11 ?VisitConcreteElementB13
?VisitConcreteElementA27 ?VisitConcreteElementB29 ?Accept13 
?AcceptA16 ?AcceptB20 ?VisitConcreteElementA24 
?VisitConcreteElementB26 ?ConcreteVisitor15 ?ConcreteVisitor211 
?Element14 ?ConcreteElementA18 ?ConcreteElementB22 
WHERE {
?Visitor27  a woc:Interface .
?VisitConcreteElementA11  a woc:Method .
?VisitConcreteElementB13  a woc:Method .
?VisitConcreteElementA27  a woc:Method .
?VisitConcreteElementB29  a woc:Method .
?Accept13  a woc:Method .
?AcceptA16  a woc:Method .
?AcceptB20  a woc:Method .
?VisitConcreteElementA24  a woc:Method .
?VisitConcreteElementB26  a woc:Method .
?ConcreteVisitor15  a woc:Class .
?ConcreteVisitor211  a woc:Class .
?Element14  a woc:Class .
?Element14 woc:hasModifier woc:Abstract .
?ConcreteElementA18  a woc:Class .
?ConcreteElementB22  a woc:Class .
?ConcreteVisitor15 woc:hasMethod ?VisitConcreteElementA11 .
?VisitConcreteElementA11 woc:hasParameter ?cA10 .
?cA10 woc:hasType ?ConcreteElementA18 .
?ConcreteVisitor15 woc:hasMethod ?VisitConcreteElementB13 .
?VisitConcreteElementB13 woc:hasParameter ?cB12 .
?cB12 woc:hasType ?ConcreteElementB22 .
?ConcreteVisitor211 woc:hasMethod ?VisitConcreteElementA27 .
?VisitConcreteElementA27 woc:hasParameter ?cA26 .
?cA26 woc:hasType ?ConcreteElementA18 .
?ConcreteVisitor211 woc:hasMethod ?VisitConcreteElementB29 .
?VisitConcreteElementB29 woc:hasParameter ?cB28 .
?cB28 woc:hasType ?ConcreteElementB22 .
?Element14 woc:hasMethod ?Accept13 .
?Accept13 woc:hasParameter ?v12 .
?ConcreteElementA18 woc:hasMethod ?AcceptA16 .
?AcceptA16 woc:hasParameter ?vA15 .
?ConcreteElementB22 woc:hasMethod ?AcceptB20 .
?AcceptB20 woc:hasParameter ?vA19 .
?Visitor27 woc:hasMethod ?VisitConcreteElementA24 .
?VisitConcreteElementA24 woc:hasParameter ?cA23 .
?cA23 woc:hasType ?ConcreteElementA18 .
?Visitor27 woc:hasMethod ?VisitConcreteElementB26 .
?VisitConcreteElementB26 woc:hasParameter ?cB25 .
?cB25 woc:hasType ?ConcreteElementB22 .
?ConcreteElementB22 woc:extends ?Element14 .
?ConcreteElementA18 woc:extends ?Element14 .
?ConcreteVisitor15 woc:implements ?Visitor27 .
?ConcreteVisitor211 woc:implements ?Visitor27 .
?AcceptA16 woc:references ?VisitConcreteElementA24 .
?AcceptB20 woc:references ?VisitConcreteElementB26 .
}
		\end{Verbatim}
	\end{minipage}
\end{lrbox}

\resizebox{0.6\columnwidth}{!}{\usebox\myvc}

It can be seen in the above query that Model2Mine also captures behavioral features of the pattern. For instance, the main characteristic of a Visitor pattern is that an Element class will have an Accept method which has a parameter of Visitor type. In the body of the accept method, the visitor object calls the respective visit method for the element. The above query is granular enough to clearly describe the parameter types of the Accept methods as well as the Visit methods (lines 27,30,33,36,45,48). It also captures the reference of visit methods within the accept method (lines 53-54). 

\section{Discussion}\label{Discussion}
Currently Model2Mine handles components like class, interface and methods as well as relationships like generalization, association and interface realization. This section covers threats to validity as well as current limitations of the tool.

\subsection{Threats to Validity}
Internal Validity: The experiments conducted to evaluate accuracy of the model used random projects available on the Internet.  Repeated attempts at running a generated query does not alter the results. However, due to time and system memory constraints, queries were executed and validated only on projects with lines of code in the range of 2000-5000. 

External Validity: There may be limited generalizability, as projects included in this study might not have necessary variation in implementation of the code that will cause a deterioration in the performance of the queries. For instance, while we observed a high precision and recall for patterns that can be uniquely identified using stereotypes and other behaviors that can be represented through filters or comments, there could be projects where classes unrelated to patterns have similar features with classes that belong to design patterns.  Additional projects need to be examined to address this issue. 
\subsection{Limitation: Multi-language Support}\label{Limitation1}
For some of the larger open source projects that were analysed, when a class is inherited from a C++ library using .h headers into a Java project, such classes were not accounted for in the triples created by CodeOntology. Due to this constraint, open source projects that had design patterns implemented but depended on C++ libraries had to be removed from the test dataset. Only if a multi-language environment is used to generate triples and a corresponding API is used to parse the SPARQL query, the full language-agnostic potential of Model2Mine could be utilized. 

\subsection{Limitation: Nested Classes}\label{Limitation2}
One of the challenges we identified about the use of UML diagrams to generate SPARQL queries is that, CodeOntology triples created for nested classes differ from triples created for non-nested classes. Hence a query generated using a UML diagram for a pattern might have a high false negative rate when trying to parse projects with nested classes. 

\subsection{Limitation: Granularity}\label{Limitation3}
The library still needs to be made more granular for handling collaboration and composition relationships. The FILTER section in SPARQL query is another feature that is not fully supported. For instance, filtering results based on distinction is enabled. That is if two classes are to be derived based on some relation, it can be ensured that the classes are not the same. While this feature is not required for most patterns, if implemented, this would improve accuracy for complex nested patterns. 

\subsection{Limitation: UML Ambiguities}\label{Limitation4}
In our approach, design patterns are described using UML (Unified Modeling Language), which is semi-formal in nature. The UML notation may lead to ambiguities and inconsistencies \cite{b4}. The accuracy of our results can be improved if the UML diagrams used for query generations incorporate additional stereotypes, tagged values, constraints and meta model elements discussed in \cite{b22} \cite{b23}. For preliminary validation, support for stereotypes related to constructors, getters and setters, and overriding of methods was implemented. More stereotypes and elements like comments that are not currently supported by XMI conversion with traditionally used metamodels can be incorporated for increased accuracy. 

\subsection{Limitation: Shorthand queries}\label{Limitation5}
Another limitation of our current implementation is related to shorthand queries. Usually SPARQL queries are written more elegantly using shorthand and indentation to avoid redundancy and reduce the character length of the query. Due to the generic nature of the library implementation and the iterative method in which we construct the query, the query relies on SPARQL to re-order constraints.

\subsection{Limitation: Differentiating hard and soft constraints}\label{Limitation6}
The current implementation of Model2Mine does not differentiate hard and soft constraints. For example, in an implementation of Factory pattern, the UML Class Diagram shows two classes that implement the factory interface to represent the context that there are more than one implementations. However, the hard constraint is only to have at least one implementation. While the current implementation does not automatically identify this, intelligent ways can be incorporated to prioritize hard and soft rules. For example, WHERE statement of the query can be dedicated for hard constraints with the FILTER section handling all soft constraints. While hard constraints use "intersection" operation to filter results, soft constraints can use "union" operation. Results could further be ranked based on how many soft constraints are matched. 

\section{Conclusion} \label{Conclusion}

Model2Mine generates SPARQL query to search for a given design pattern in source code by parsing its UML diagram. It is capable of generating queries for complex design patterns like the visitor pattern. It does not limit the number of entities or relationships to be queried. The query will have the same granularity as the input UML diagram in terms of capturing structural, creational and behavioral aspects. Thus, it is able to mine more types of patterns than other techniques.

We assessed our techniques using representative patterns from the three types of design patterns: creational, structural, and behavioral.  Our feature coverage indicates that it is capable of uniquely identifying patterns by enabling stereotypes and parsing highly granular diagrams. Our initial experiments show an average precision 85.71\% and average recall of 82.29\%.  It was observed that, for patterns that can be better identified using unique stereotypes, precision and recall can be improved up to 100\%.  In addition, we also offered ways to improve accuracy when mining design patterns and lessons learned.
Future work includes the following.  Combining the capability of this model with the approach of \cite{b8}, accurate code snippet recommendations can be created. In addition, we plan to identify hard and soft constraints for each design pattern, to improve the accuracy of the tool.



\section*{Acknowledgment}

The authors wish to thank Namita Dave for her assistance with third party tools and setting up the development environment.  We also thank Elif Hepateskan for her assistance with performing evaluations.  

\end{document}